# The Closer, the Better: Engineering RGB Thermometric Performance in $Tb^{3+}$, $Eu^{3+}$-Doped MOFs

**M. Szymczak[1], L. Giang[2*], L. Marciniak[1*]**

[1] Institute of Low Temperature and Structure Research, Polish Academy of Sciences,

Okólna 2, 50-422 Wrocław, Poland

[2] Institute of Materials Science, Vietnam Academy of Science and Technology, Hanoi, Viet Nam

*corresponding author: l.marciniak@intibs.pl, lamgianghp@gmail.com

## Abstract

The effectiveness of luminescence thermometry, which exploits temperature-induced changes in the spectroscopic properties of phosphor materials for temperature sensing, has been extensively demonstrated. However, from a practical perspective, monitoring temperature-induced variations in the color of the emitted light offers a considerably more straightforward and user-friendly alternative to conventional spectral analysis. In this context, establishing a clear relationship between the thermal evolution of the emission color and the structure of the phosphor is essential for the rational design of visual luminescence thermometers. In this work, we investigate the correlation between the temperature-dependent emission color and the structural characteristics of $Tb^{3+}$, $Eu^{3+}$-co-doped metal-organic frameworks (MOFs), with particular emphasis on their potential for filter-free, camera-based thermal sensing. The obtained results demonstrate that, within the investigated series of materials, decreasing the interionic distance between $Eu^{3+}$ and $Tb^{3+}$ ions increases the probability of energy transfer and consequently enhances the thermal sensitivity of the camera-based readout derived from the ratio of luminescence intensities recorded in the blue and red channels. These findings establish a direct relationship between the structural characteristics of the luminescent material, the underlying energy-transfer processes, and its thermometric performance. More broadly, this study provides a foundation for the rational design of visual luminescence thermometers with predefined sensing characteristics, opening a pathway toward structure-guided optimization of filter-free, camera-based thermal sensing platforms.

## Introduction

Although thermally induced changes in the spectroscopic properties of phosphors are widely exploited in luminescence thermometry, the vast majority of reported studies are based on the ratiometric approach. This is fully justified, as the intensity ratio of two emission bands is generally regarded as one of the most reliable spectroscopic parameters, being only minimally affected by factors such as the detection geometry, excitation power density, or the concentration and spatial distribution of the phosphor on the investigated surface. The effectiveness of this approach has been demonstrated in numerous luminescent thermometric systems. In most cases, however, the temperature readout relies on the intensity ratio of emission bands located in close spectral proximity. While this strategy offers several advantages, it typically requires the acquisition of emission spectra using a spectrometer, which increases both the complexity and cost of the measurement setup. An attractive alternative is the analysis of thermally induced changes in the emission color. When emission bands of the luminescent thermometer exhibiting different thermal responses are spectrally well separated, temperature variations may lead to noticeable changes in the color of the emitted light. Such an approach not only enables direct visual observation of temperature changes by the naked eye but can also be employed for accurate thermal mapping using a conventional digital camera. From this perspective, $Eu^{3+}$ and $Tb^{3+}$ ions are particularly attractive luminescent centers owing to their well-defined red and green emissions, respectively. Furthermore, the energy difference between their dominant emitting excited states, namely the $^5D_4$ level of $Tb^{3+}$ and the $^5D_0$ level of $Eu^{3+}$, is sufficiently large that energy transfer between them requires phonon assistance. Such phonon-assisted energy-transfer processes are highly sensitive to temperature and therefore constitute an excellent basis for luminescence thermometry. However, a limitation of lanthanide-based luminescent thermometers is the relatively low absorption cross-section of the 4*f*-4*f* transitions, which results in modest luminescence brightness. To overcome this drawback, lanthanide ions are frequently incorporated into metal-organic frameworks (MOFs), where the organic ligand acts as an efficient sensitizer of lanthanide luminescence. This strategy not only enhances the absorption of excitation light due to the much larger absorption cross-section of the ligand electronic transitions, but also provides additional opportunities for tuning thermometric performance of such a sensor. In these systems, excitation energy is initially absorbed by the ligand causing a transition of the electron from the ground to the singlet state, followed by intersystem crossing to the triplet state, and subsequent energy transfer to the lanthanide ions. Consequently, the energy position of the triplet state can significantly influence the thermometric properties of the material. Moreover, the structure of the MOF determines the average distances between lanthanide ions, thereby affecting the efficiency of the interionic energy-transfer processes between them. Achieving precise control over the thermometric performance of $Tb^{3+}$, $Eu^{3+}$ co-doped systems therefore requires a comprehensive understanding of how the MOF host influences the thermal stability and luminescence behavior of these ions.

In the present work, we systematically investigate the effect of different MOF hosts on the thermometric properties of $Tb^{3+}$, $Eu^{3+}$-based luminescent thermometers. To this end, the temperature-dependent spectroscopic properties of Gd-BTC:$Tb^{3+}$, $Eu^{3+}$, Zr-BTC:$Tb^{3+}$, $Eu^{3+}$, Zr-BDC:$Tb^{3+}$, $Eu^{3+}$ and Zr-TA:$Tb^{3+}$, $Eu^{3+}$ were examined. The results provide valuable insights into the relationship between MOF structure and thermometric performance, offering guidelines for the rational design of visual luminescent thermometers with properties tailored to the requirements of specific applications.

**Experimental Section**

*Materials*

All chemicals are were purchased from Sigma Aldrich and used without further purification: Gadolinium(III) nitrate hexahydrate ($Gd(NO_3)_3·6H_2O$, 99.99%), Zirconyl chloride octahydrate ($ZrOCl_2·8H_2O$, 98%), Europium(III) chloride hexahydrate ($EuCl_3·6H_2O$, 99.99%), Terbium(III) chloride hexahydrate ($TbCl_3·6H_2O$, 99.9%), Europium(III) nitrate pentahydrate ($Eu(NO_3)_3·5H_2O$, 99.99%), Terbium(III) nitrate pentahydrate ($Tb(NO_3)_3·5H_2O$ , 99.99%) Trimesic acid ($H_3BTC$, 95%), Terephthalic acid ($H_2BDC$, 98%), L-(+)-Tartaric acid ($H_2TA$, 99.7%), N,N-Dimethylformamide (DMF, 99%), Ethyl alcohol (EtOH, ≥95%), Diethylene glycol (DEG, 99 %), Methanol (MeOH, 99%), N,N-Dimethylformamide (DMF, 99,8%).

*Synthesis*

The MOF samples were synthesized using a hydrothermal or solvothermal method. Appropriate amounts of $Gd^{3+}$, $Zr^{4+}$, $Eu^{3+}$, and/or $Tb^{3+}$ precursor salts, depending on the intended composition of the particular MOF, were dissolved in the corresponding solvent or solvent mixture and stirred magnetically for 30 min to obtain a homogeneous metal precursor solution. The detailed amounts and concentrations of the metal precursors used for each sample are summarized in Table 1, below. Simultaneously, the appropriate organic ligand, namely $H_3BTC$, $H_2BDC$ or $H_2TA$, was dispersed or dissolved in the corresponding solvent system. The ligand amounts and solvent compositions used for the synthesis of the individual samples are also provided in Table 1. The ligand solution was subjected to ultrasonication for 15-30 min and subsequently stirred for an additional 15 min to obtain a homogeneous organic precursor solution. The resulting ligand precursor was then added to the previously prepared metal salt solution, and the mixture was stirred at room temperature until a homogeneous suspension was obtained. The resulting reaction mixtures were transferred into a Teflon-lined stainless-steel autoclave, tightly sealed, and subjected to hydrothermal or solvothermal treatment for 24 h. The reaction temperature was set to 393 K for the $Eu^{3+}$- and $Tb^{3+}$-BTC, Gd-BTC:$Eu^{3+}$, Gd-BTC:$Tb^{3+}$, Gd-BTC:$Eu^{3+}$,$Tb^{3+}$ and Zr-TA:$Eu^{3+}$,$Tb^{3+}$ samples, whereas Zr-BDC:$Eu^{3+}$,$Tb^{3+}$ and Zr-BTC:$Eu^{3+}$,$Tb^{3+}$ were synthesized at 473 K. After completion of the reaction, the products were collected by filtration and purified using the procedure appropriate for the respective MOF. The $Tb^{3+}$-BTC, Gd-BTC:$Eu^{3+}$, Gd-BTC:$Tb^{3+}$, Gd-BTC:$Eu^{3+}$,$Tb^{3+}$ samples were redispersed in DMF, thoroughly mixed, and soaked for 24 h, followed by three washing cycles with methanol using centrifugation at 5400 rpm for 10 min. The Zr-BDC:$Eu^{3+}$,$Tb^{3+}$ and Zr-BTC:$Eu^{3+}$,$Tb^{3+}$ products were centrifuged once with deionized water, immersed in 50 mL of DMF overnight, and subsequently washed twice with methanol by centrifugation. The Zr-TA:$Eu^{3+}$,$Tb^{3+}$ products were washed three times with methanol by centrifugation. Finally, all samples were dried at 333 K. Following these procedures, $Tb^{3+}$-BTC, $Eu^{3+}$-BTC, Gd-BTC:5%$Eu^{3+}$, Gd-BTC:20%$Tb^{3+}$, Gd-BTC:5%$Eu^{3+}$,20%$Tb^{3+}$ (Gd-BTC), Zr-BDC:5%$Eu^{3+}$,20%$Tb^{3+}$ (Zr-BDC), Zr-BTC:5%$Eu^{3+}$,20%$Tb^{3+}$ (Zr-BTC), and Zr-TA:5%$Eu^{3+}$ (Zr-TA), 20%$Tb^{3+}$ MOFs were obtained.

**Table 1**. Amounts of reagents used for the synthesis of the investigated MOFs, including the metal precursors, organic ligands, and solvents.

<table>
<tr><th></th><th colspan="4">Metal ions</th><th></th><th></th></tr>
<tr><th></th><th>$Gd^{3+}$ source (0.25M)</th><th>$Zr^{4+}$ source (0.25M)</th><th>$Eu^{3+}$ source (0.25M)</th><th>$Tb^{3+}$ source (0.25M)</th><th>DMF (ml)</th><th>Organic ligand</th></tr>
<tr><th>Tb-BTC</th><td>-</td><td rowspan="5">-</td><td></td><td>20 ml</td><td rowspan="5">-</td><td rowspan="5">2 g of $H_3BTC$ (20 ml of DMF, 30 ml of MeOH)</td></tr>
<tr><th>Eu-BTC</th><td>-</td><td>20 ml</td><td></td></tr>
<tr><th>Gd-BTC:$Tb^{3+}$</th><td>17.79 ml</td><td></td><td>2.21 ml</td></tr>
<tr><th>Gd-BTC:$Eu^{3+}$</th><td>14.49 ml</td><td>0.51 ml</td><td></td></tr>
<tr><th>Gd-BTC:$Eu^{3+}$,$Tb^{3+}$</th><td>17.15 ml</td><td>0.57 ml</td><td>2.28 ml</td></tr>
</table>

<table>
<tr><td>Zr-BTC:$Eu^{3+}$,$Tb^{3+}$</td><td rowspan="3">-</td><td rowspan="3">15 ml</td><td rowspan="3">1 ml</td><td rowspan="3">4 ml</td><td rowspan="2">10</td><td>3.5 g of $H_3BTC$ (30 ml of MeOH)</td></tr>
<tr><td>Zr-BDC:$Eu^{3+}$,$Tb^{3+}$</td><td>1 g of $H_2BDC$ (30 ml of MeOH)</td></tr>
<tr><td>Zr-TA:$Eu^{3+}$,$Tb^{3+}$</td><td>-</td><td>1.35 g of $H_2TA$ (10 ml of DMF)</td></tr>
</table>

*Methods*

The luminescence measurements including emission, excitation spectra and luminescence kinetics were performed with a FLS1000 Fluorescence Spectrometer from Edinburgh Instruments, equipped with a 450 W xenon lamp and μFlash pulsed lamp as an excitation source and an R928 photomultiplier tube from Hamamatsu as a detector. The temperature-dependent measurements were performed using a THMS 600 heating-cooling stage from Linkam (temperature stability of 0.1 K and a set point resolution of 0.1 K). Before each measurement, the temperature was stabilized for 1 min to ensure reliable readouts. The morphology of synthesized Zr-BDC MOF was examined using a scanning electron microscope (SEM, Ultra Plus, Zeiss, Oberkochen, Germany) operated at 25 kV. The system was equipped with an energy-dispersive X-ray spectrometer (EDS, EMAX ENERGY) featuring an XFlash 5010 detector with a thickness of 0.45 mm operated at 15 keV (Bruker Nano GmbH, Berlin, Germany). The structural analysis was performed using a PANalytical X'Pert Pro powder diffractometer system with CuKα radiation ($\lambda$ = 1.54060 Å) in the 2θ range of 5°–100°. The Fourier transform infrared spectroscopy (FTIR) analysis of the samples was conducted on a Fourier Transform Infrared Spectrometer, Bruker, model Tensor 27. The DSC-TG analysis was conducted on a simultaneous thermal analyzer, model STA 449 F1 Jupiter (NETZSCH).

**Results and discussion**

The morphology of $Tb^{3+}$ and $Eu^{3+}$ co-doped Gd-BTC and the Zr-based MOFs synthesized using BTC, BDC and TA as organic ligands was investigated by scanning electron microscopy.

As shown in Figures 1a-d (also in SI on Figures S1-S3), all synthesized materials exhibit a well-defined particulate morphology; however, pronounced differences in particle shape and size are observed depending on the composition and synthesis conditions. In the case of the BTC-based materials, the morphology differs markedly between Gd-BTC and Zr-BTC. Gd-BTC consists of relatively uniform and well-defined elongated tubular-like particles with lengths on the order of several hundred nanometers. In contrast, Zr-BTC is composed of irregularly shaped particles with a broad size distribution, which tend to form larger agglomerates. At this stage, it is difficult to unambiguously determine whether these morphological differences originate primarily from the nature of the metal cation or from the different synthesis conditions. In particular, Zr-BTC was synthesized at a considerably higher temperature (473 K) than Gd-BTC (393 K), which may strongly affect nucleation, crystal growth, and particle aggregation.[1] The possible influence of the synthesis temperature is further supported by the morphology of Zr-BDC, which was also prepared at elevated temperature and exhibits features similar to those observed for Zr-BTC, including irregular particle shapes and a relatively broad particle-size distribution. A distinctly different morphology was observed for the Zr-TA samples. Both materials consist predominantly of aggregated nanoparticles with a considerably narrower size distribution. The individual nanoparticles are approximately 50-100 nm in size and form porous, loosely packed agglomerates. Such morphology is markedly different from that of the Zr-BTC and Zr-BDC samples and indicates that the nature of the organic ligand, together with the corresponding synthesis conditions, strongly affects particle nucleation and growth. The elemental composition of the synthesized samples was additionally investigated by EDX spectroscopy. Figure 1d shows a representative SEM image of the Zr-TA sample with the region selected for EDX analysis, while the corresponding spectrum is presented in Figure 1e. EDX data for the remaining samples are provided in Figures S4-S5. For the Zr-TA sample, the analysis confirmed the presence of Zr, with an atomic content of approximately 7.04%, together with substantial amounts of C and O, corresponding to 38.74 and 52.28 at.%, respectively. The simultaneous presence of Zr, C, and O is consistent with the expected composition of a Zr-based material containing tartaric-acid-derived organic components.

The XRD patterns of the investigated samples are presented in Figure S6. The diffraction pattern of Gd-BTC is consistent with those previously reported for analogous Gd-BTC-based frameworks. In contrast, all Zr-based samples exhibit diffraction profiles markedly different from those typically observed for highly crystalline Zr-MOFs such as UiO-66 or MOF-808. Instead of sharp and intense Bragg reflections characteristic of long-range periodic ordering, the patterns are dominated by a broad amorphous halo accompanied by several broadened reflections centered at approximately $2\theta = 29.76°$, 34.17°, 49.86°, and 58.98°. These reflections can be indexed to the (101), (002)/(110), (112)/(200), and (211) planes of metastable tetragonal/cubic $ZrO_2$, respectively.[2] The severe peak broadening (large FWHM) directly reflects spatial confinement effects, confirming the in-situ generation of sub-5 nm t/c-$ZrO_2$ nanoclusters possessing only short-range structural order.[3] The formation of this hybrid architecture-comprising an amorphous MOF matrix embedded with t/c-$ZrO_2$ nanoclusters (aMOF@t/c-$ZrO_2$) - originates from the localized self-assembly and partial condensation of [$Zr_6O_4(OH)_4$] secondary building units (SBUs) under coordinatively incomplete or linker-deficient synthetic conditions.[2,4] Rather than constituting a structural drawback, the amorphous-nanocluster architecture may offer several advantages over fully crystalline counterparts. The intrinsic disorder of the amorphous MOF matrix, combined with missing-linker defects, creates a high density of coordinatively unsaturated $Zr^{4+}$ Lewis acid sites and Zr-OH Brønsted acid sites.[4,5] Simultaneously, the loss of strict long-range periodicity can generate a hierarchical micro-/mesoporous network that reduces steric constraints and promotes more efficient mass transport, especially for bulky species.[6,7] Moreover, the in situ generated $ZrO_2$ nanoclusters can serve as rigid reinforcing domains, improving the mechanical and hydrolytic stability of the material and helping to preserve its structural integrity under demanding conditions.[8]

To obtain further information on the chemical bonding and local coordination environment within the Zr-based materials, FTIR spectroscopy was subsequently performed. The FTIR spectra of the investigated samples are shown in Figure 1f. The FTIR spectrum of Zr-TA displays a different set of characteristic vibrations, consistent with the presence of tartaric-acid-derived ligands. A broad O-H stretching band is observed around 3300 $cm^{-1}$ indicating a smaller contribution from adsorbed water or residual solvent. The most prominent spectral features are the intense band at approximately 1587 $cm^{-1}$ and a shoulder near 1653 $cm^{-1}$, which can be assigned to asymmetric stretching vibrations of coordinated carboxylate groups. Corresponding symmetric carboxylate stretching modes are observed at approximately 1415 and 1379 $cm^{-1}$. The bands at approximately 1109 and 1062 $cm^{-1}$ are associated with C-O stretching vibrations of the secondary hydroxyl groups within the tartaric acid backbone, whereas weaker features near 1303, 1252, and 1243 $cm^{-1}$ can be attributed to C-H and O-H deformation modes.[10] Collectively, these results are consistent with deprotonation of the carboxylic groups of tartaric acid and their coordination to Zr centers. Similarly, the FTIR spectra of the Zr-BTC and Zr-BDC samples show the characteristic vibrations of carboxylate-containing aromatic ligands. In both cases, the disappearance or pronounced reduction of the free carboxylic acid C=O stretching band, together with the appearance of distinct asymmetric and symmetric $COO^{-}$ stretching bands, supports deprotonation of the carboxyl groups and their interaction with Zr-containing inorganic nodes. Additional bands in the fingerprint region can be assigned to vibrations of the aromatic rings and C-O bonds of the BTC and BDC ligands.[11,12] Thus, despite the low degree of long-range crystallinity observed by PXRD, the FTIR spectra confirm the presence of intact organic ligands coordinated within the Zr-containing materials.

The thermal behavior of the Zr-BDC and Zr-BTC samples was additionally examined by simultaneous thermogravimetric and differential scanning calorimetry analysis (Figure S7 and S8). For the Zr-BDC (Figure S7) the TG curve reveals an initial mass loss of approximately 12.22% between 300 and 630 K. This region is accompanied by two thermal events in the DSC profile: a broad endothermic signal centered at approximately 362 K and a much weaker exothermic feature near 587 K. These events are attributed mainly to the gradual removal of physically adsorbed water and residual solvent molecules, possibly accompanied by rearrangements within the local coordination environment. Above approximately 630 K, a second mass-loss region is observed, corresponding to an additional decrease in mass of about 8.2%. This process coincides with a broad and intense exothermic DSC signal extending over approximately 708-853 K, with an enthalpy change of $\Delta H \approx 276.7$ J $g^{-1}$. The onset and offset temperatures of this event are approximately 708 K and 853 K, respectively. The close correlation between the TG and DSC profiles indicates that this temperature interval corresponds to the major decomposition of the organic BDC-containing component and the associated collapse or transformation of the Zr-organic structure. Above approximately 853 K, only minor additional mass changes are observed, indicating that the organic fraction has been largely removed and that the remaining material consists predominantly of a thermally stable inorganic Zr-containing residue, most plausibly zirconia-based. Accordingly, the TG/DSC results indicate that the Zr-BDC material retains its organic framework components up to approximately 700 K, while extensive thermal decomposition occurs mainly within the 708-853 K range. A similar thermal behavior was observed for Zr-BTC, although the corresponding mass losses were lower, amounting to approximately 6.95% in the low-temperature region and 5.94% during the main decomposition step. The TG/DSC curves for Zr-BTC are provided in the Supporting Information (Figure S8). The reduced mass losses observed for Zr-BTC may indicate a lower content of retained volatile species and thermally decomposable organic components, or a higher relative contribution of the inorganic Zr-containing fraction compared with Zr-BDC.

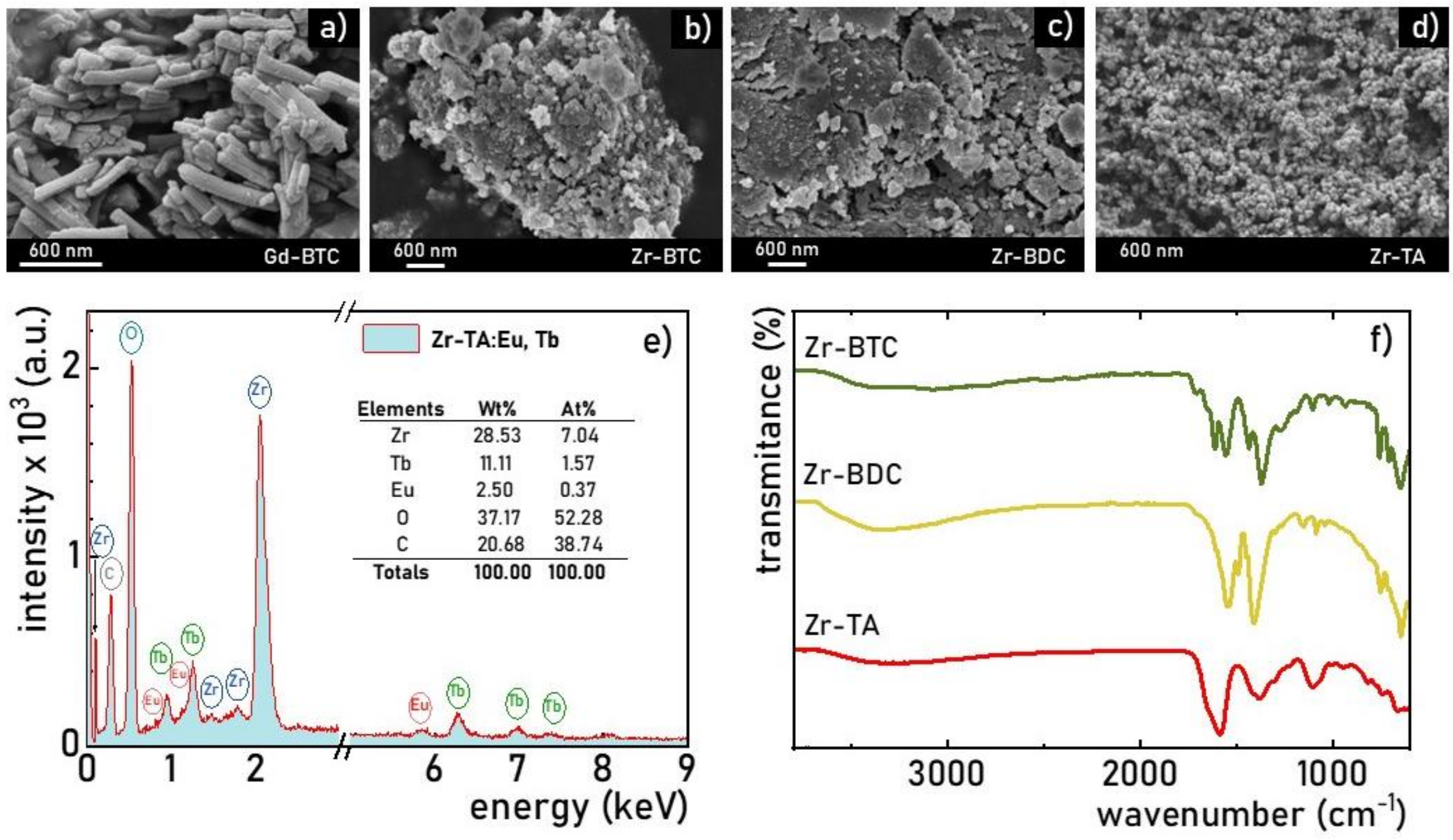

**Figure 1.** Representative SEM images of: Gd-BTC -a), Zr-BTC -b), Zr-BDC -c) and Zr-TA -d); the representative EDX spectrum of Zr-TA -e). FTIR spectra of the investigated Zr-based MOFs -f).

Since the aim of this work is to develop a luminescent thermometer based on the luminescence intensity ratio of $Eu^{3+}$ and $Tb^{3+}$ ions, it is essential to investigate a system co-doped with both activators. However, before that the understanding of the spectroscopic of complexes solely doped with single type of ion is of fundamental significance. Therefore, in this section, the luminescence properties of Eu-BTC and Tb-BTC are compared with those of their corresponding Gd-based analogues, Gd-BTC:5%$Eu^{3+}$ and Gd-BTC:20%$Tb^{3+}$. The motivation for replacing $Eu^{3+}$ and $Tb^{3+}$ host ions with $Gd^{3+}$ originates from the fact that $Gd^{3+}$ does not exhibit intense emission bands within the visible spectral range.[13] Consequently, the presence of $Gd^{3+}$ does not interfere with the optical temperature readout while providing an optically inactive host matrix for the emitting lanthanide ions. The spectroscopic properties of lanthanide complexes originate from the characteristic intra-configurational 4*f*-4*f* electronic transitions, which give rise to the narrow emission lines observed in their luminescence spectra. In contrast, in these systems the excitation process is predominantly mediated by the organic ligand through the well-known antenna effect.[14,15] Upon excitation at an appropriate wavelength, electrons are promoted from the ligand ground state (*GS*) to the excited singlet state (*S*) (Figure 2a). Subsequently, intersystem crossing occurs, resulting in population of the energetically lower triplet state (*T*). From this level, energy is transferred to the excited states of the lanthanide ions, leading to the occurrence their characteristic emission bands in the spectra. Since the absorption cross-section of the ligand is several orders of magnitude larger than that of the parity-forbidden 4*f*-4*f* transitions of lanthanide ions, excitation through the ligand provides a much more efficient population pathway for the emitting states, resulting in significantly enhanced luminescence intensity compared to direct excitation of the lanthanide ions.[15,16] This mechanism is confirmed by the excitation spectra recorded while monitoring the characteristic emissions of $Eu^{3+}$ and $Tb^{3+}$ ions (Figure 2b). In addition to the narrow excitation bands corresponding to the 4*f*-4*f* transitions of the lanthanide ions, a broad and intense excitation band is observed in the ultraviolet region, originating from the $GS \rightarrow S$ transition of the organic ligand. Notably, the excitation spectrum of $Eu^{3+}$ in Gd-BTC:5%$Eu^{3+}$ also exhibits an additional sharp band centered at approximately 320 nm, which is attributed to the characteristic $^8S_{7/2} \rightarrow ^6P_{5/2}$ electronic transition of $Gd^{3+}$ ions (Figure 2b). The presence of this band indicates that $Gd^{3+}$ ions may also participate in the excitation and energy transfer processes within this system. The luminescence of $Tb^{3+}$ ions originates primarily from the radiative depopulation of the $^5D_4$ excited state to the $^7F_J$ manifold (Figure 2a). Among these transitions, the $^5D_4 \rightarrow ^7F_5$ and $^5D_4 \rightarrow ^7F_6$ transitions, located at approximately 545 and 490 nm, respectively, exhibit the highest oscillator strengths and therefore dominate the emission spectra.[17,18] Additional emission bands observed at approximately 585 and 620 nm correspond to the $^5D_4 \rightarrow ^7F_4$ and $^5D_4 \rightarrow ^7F_3$ transitions, respectively. All of these emission bands can be observed in emission spectra of both Tb-BTC and Gd-BTC:20%$Tb^{3+}$, and no significant differences in either the spectral positions or relative emission intensities are observed between these compounds (Figure 2c). In contrast, $Eu^{3+}$ ions exhibit their characteristic intense red luminescence arising from radiative relaxation of the $^5D_0$ excited state to the $^7F_J$ manifold. The two dominant emission bands correspond to the $^5D_0 \rightarrow ^7F_1$ and $^5D_0 \rightarrow ^7F_2$ transitions, centered at approximately 590 and 620 nm, respectively. An important distinction between these transitions is that the $^5D_0 \rightarrow ^7F_1$ transition is a magnetic dipole transition, whose intensity is only weakly affected by the local crystal field, whereas the hypersensitive $^5D_0 \rightarrow ^7F_2$ transition is an electric dipole transition and is strongly influenced by the local symmetry surrounding the $Eu^{3+}$ ion.[19,20] Consequently, the relative intensity of these two transitions serves as a sensitive probe of the local coordination environment.[21] In addition to the dominant red emission bands, weaker emission lines located at approximately 575, 650, and 700 nm are observed and are assigned to the $^5D_0 \rightarrow ^7F_0$, $^5D_0 \rightarrow ^7F_3$, and $^5D_0 \rightarrow ^7F_4$ transitions, respectively. The observation of the $^5D_0 \rightarrow ^7F_0$ transition in both Eu-BTC and Gd-BTC:5%$Eu^{3+}$ provides direct evidence that $Eu^{3+}$ ions occupy crystallographic sites without inversion symmetry.[22–24]

An important consideration for the design of dual-emitting systems is the relative energetic positions of the excited states of $Tb^{3+}$ and $Eu^{3+}$ ions. The $^5D_4$ excited state of $Tb^{3+}$ (~20,500 $cm^{-1}$) lies at a considerably higher energy than the $^5D_0$ emitting level of $Eu^{3+}$ (~17,200 $cm^{-1}$), making phonon-assisted energy transfer from $Tb^{3+}$ to $Eu^{3+}$ energetically feasible in co-doped materials. Therefore, one may expect the occurrence of $Tb^{3+} \rightarrow Eu^{3+}$ energy transfer in systems simultaneously containing both lanthanide ions.[25,26] Furthermore, the emitting levels of both $Tb^{3+}$ ($^5D_4$) and $Eu^{3+}$ ($^5D_0$) are separated from the lower-lying electronic states by more than 10,000 $cm^{-1}$. Such a large energy gap effectively suppresses multiphonon nonradiative relaxation, thereby minimizing thermal depopulation of the emitting states and ensuring high intrinsic luminescence stability.

Since the luminescence intensity ratio between the $^5D_0 \rightarrow {}^7F_2$ and $^5D_0 \rightarrow {}^7F_1$ transitions of $Eu^{3+}$ is highly sensitive to changes in the local crystal field surrounding the $Eu^{3+}$ ion, this parameter ($LIR_1$) provides valuable insight into the structural evolution of the investigated compounds as a function of temperature:

$$LIR_1 = \frac{\int I(Eu^{3+} : {}^5D_0 \rightarrow {}^7F_2) d\lambda}{\int I(Eu^{3+} : {}^5D_0 \rightarrow {}^7F_1) d\lambda} \quad (1)$$

Analysis of the temperature dependence of $LIR_1$ reveals only minor variations within the 93-400 K temperature range for Eu-BTC, indicating excellent structural stability of this compound over the investigated temperature interval (Figure 2d). Interestingly, the initial $LIR_1$ value recorded at 93 K for Gd-BTC:5%$Eu^{3+}$ is slightly lower than that observed for Eu-BTC. Furthermore, increasing temperature results in a gradual and monotonic increase of $LIR_1$. This behavior suggests a slight reduction in the local symmetry around $Eu^{3+}$ ions with increasing temperature, which may originate from the somewhat lower rigidity of the Gd-BTC framework resulting from the coexistence of $Gd^{3+}$ and $Eu^{3+}$ ions within the structure.[22] Nevertheless, these changes remain relatively small, confirming that the local coordination environment of $Eu^{3+}$ is well preserved over the investigated temperature range. Since the primary objective of this work is to evaluate the potential of these materials for luminescence thermometry, it is essential to investigate the thermal stability of the emission originating from individual $Eu^{3+}$ and $Tb^{3+}$ centers. In the case of Eu-BTC, an unusual thermally induced enhancement of the emission intensity is observed (Figure 2e). As the temperature increases from 93 to approximately 280 K, the luminescence intensity increases by nearly 15%. Further heating results in the onset of thermal quenching, leading to a gradual decrease in emission intensity at higher temperatures. A markedly different behavior is observed for Gd-BTC:5%$Eu^{3+}$. In this material, the luminescence remains nearly constant up to approximately 300 K, above which thermal quenching becomes evident (Figure 2e). As a consequence, the emission intensity decreases progressively, reaching approximately 80% of its initial value at 373 K. The distinct thermal stability exhibited by Eu-BTC and Gd-BTC:5%$Eu^{3+}$ may be associated with the presence of $Gd^{3+}$ ions in the latter compound. This hypothesis is supported by the appearance of the sharp excitation line at approximately 320 nm in the excitation spectrum of Gd-BTC:5%$Eu^{3+}$, which originates from the characteristic electronic transition of $Gd^{3+}$ ions. The presence of this additional excitation pathway suggests that energy transfer to $Eu^{3+}$ does not occur exclusively through the conventional ligand-mediated mechanism involving the triplet state of the organic linker. Instead, an alternative pathway involving $Gd^{3+}$ ions may contribute to the excitation process, following the sequence $T \rightarrow Gd^{3+} \rightarrow {}^5D_0$ ($Eu^{3+}$).[27,28] Under these conditions, the thermal stability of $Eu^{3+}$ luminescence becomes partially governed by the thermal stability of the intermediate excited state of $Gd^{3+}$. Although this excited state is well separated from the $Gd^{3+}$ ground state, it is energetically located in the vicinity of several excited states of $Eu^{3+}$. Consequently, thermally activated interactions between these closely spaced energy levels may introduce additional nonradiative relaxation channels, thereby reducing the overall thermal stability of $Eu^{3+}$ emission in Gd-BTC:5%$Eu^{3+}$ compared with the parent Eu-BTC compound. The observed differences clearly demonstrate that replacing $Eu^{3+}$ host ions with optically inactive $Gd^{3+}$ ions does not simply dilute the emitting centers but also modifies the excitation and energy transfer mechanisms responsible for populating the $Eu^{3+}$ emitting state. As a result, the thermal response of the luminescence is altered despite the nearly identical emission spectra exhibited by both compounds. These observations underline the importance of carefully considering the role of the host lattice and energy migration pathways when designing lanthanide-based materials for optical thermometry applications. In contrast to the $Eu^{3+}$-activated compounds, the $Tb^{3+}$-based materials exhibit a different thermal evolution of their luminescence intensity. For Tb-BTC, a gradual decrease in emission intensity is observed over the entire investigated temperature range, with the luminescence intensity decreasing continuously to approximately 50% of its initial value at 413 K (Figure 2f). This behavior indicates the progressive activation of non-radiative relaxation processes with increasing temperature. A significantly different thermal response is observed for Gd-BTC:20%$Tb^{3+}$. In this compound, the luminescence intensity remains nearly constant up to approximately 300 K, demonstrating excellent thermal stability within this temperature range. Above 300 K, thermal quenching becomes apparent, leading to a gradual reduction of the emission intensity. Nevertheless, at 373 K the luminescence still retains approximately 80% of its initial intensity, indicating substantially higher thermal stability than that observed for Tb-BTC. The enhanced thermal stability of $Tb^{3+}$ luminescence in the Gd-based host suggests that the substitution of $Tb^{3+}$ by optically inactive $Gd^{3+}$ ions modifies the energy migration pathways within the framework, thereby reducing the probability of thermally activated non-radiative relaxation.[29] Unlike the $Eu^{3+}$-activated system, where the presence of $Gd^{3+}$ introduces an additional energy transfer pathway that may adversely affect the thermal stability of the emission, the Gd-BTC:20%$Tb^{3+}$ compound exhibits improved resistance to thermal quenching. This observation demonstrates that the influence of $Gd^{3+}$ on the luminescence properties strongly depends on the nature of the emitting lanthanide ion and the corresponding excitation and relaxation mechanisms.[30,31] It is worth emphasizing that both Gd-BTC compounds individually doped with $Eu^{3+}$ and $Tb^{3+}$ exhibit good thermal stability of their luminescence over a broad temperature range. This behavior is particularly advantageous for the development of luminescence thermometers based on co-doped systems, as reliable temperature readout requires stable emission from both emitting centers. The high thermal stability observed in Gd-BTC:5%$Eu^{3+}$ below approximately 300 K and in Gd-BTC:20%$Tb^{3+}$ over an even broader temperature range indicates that the Gd-BTC framework provides a suitable host matrix for the incorporation of both lanthanide ions.

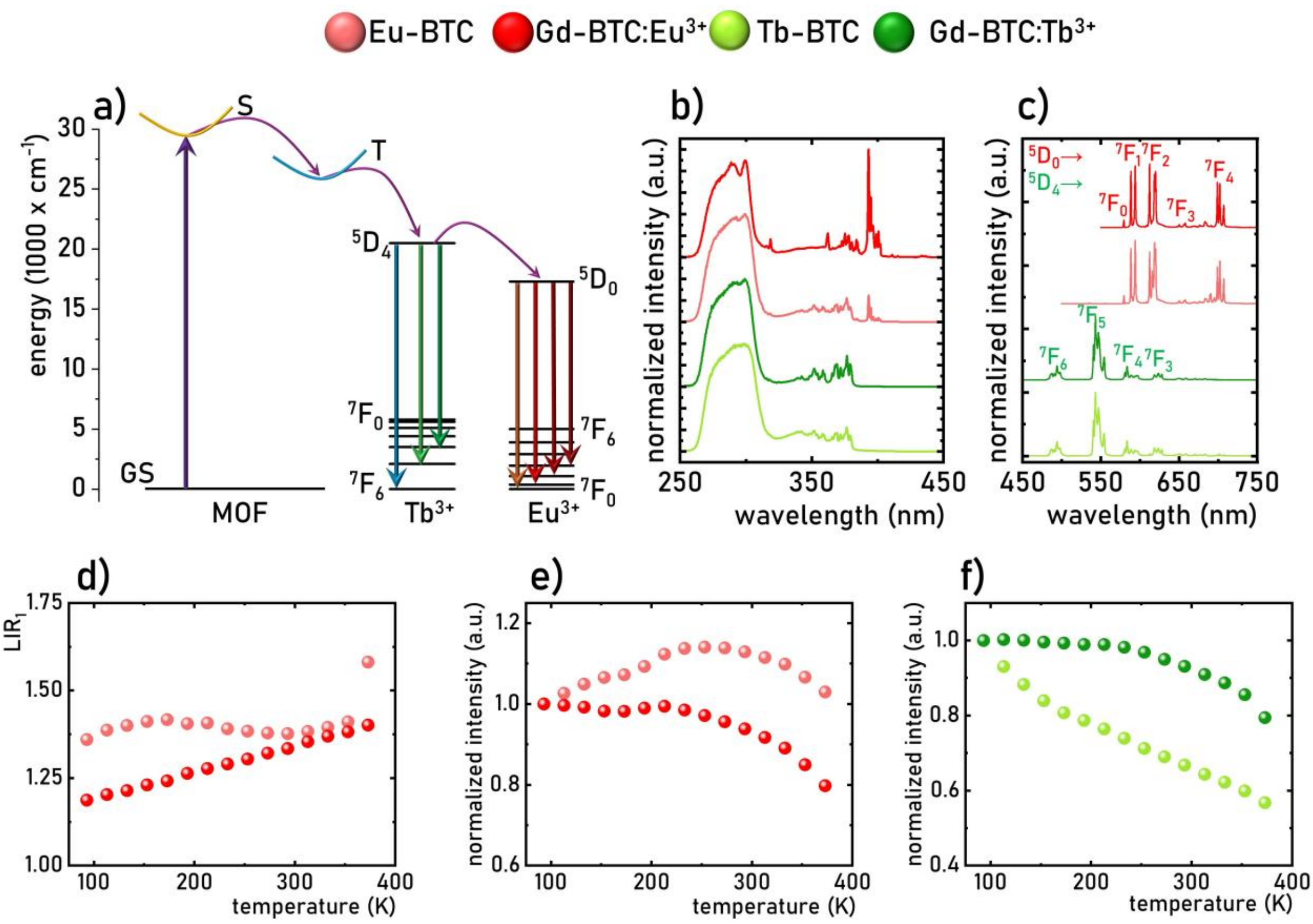

**Figure 2.** Simplified energy level diagram of $Eu^{3+}$ and $Tb^{3+}$ ions with singlet and triplets states of MOF structure – a), comparison of the excitation – b) and emission – c) spectra of Tb-BTC, Eu-BTC, Gd-BTC:$Tb^{3+}$, Gd-BTC: $Eu^{3+}$; the influence of temperature on $LIR_1$ for Eu-BTC and Gd-BTC: $Eu^{3+}$ -d); thermal dependence of integrated emission intensities of $Eu^{3+}$ ions in Eu-BTC and Gd-BTC: $Eu^{3+}$ - e) and $Tb^{3+}$ ions in Tb-BTC and Gd-BTC: $Tb^{3+}$ - f).

To evaluate the applicability of Gd-BTC:$Tb^{3+}$, $Eu^{3+}$ for luminescence thermometry, temperature-dependent emission spectra were recorded in the 83-413 K temperature range (Figure 3a). The obtained emission spectra exhibit a number of thermally induced changes. Most notably, the integrated luminescence intensity decreases by approximately 60% over the investigated temperature range. This behavior results from multiphonon nonradiative depopulation of the excited states of both $Tb^{3+}$ and $Eu^{3+}$ ions. A detailed analysis of the emission spectra reveals that the luminescence intensity of $Tb^{3+}$ ions decreases significantly faster than that of $Eu^{3+}$ ions. This effect originates from a phonon-assisted energy transfer process between $Tb^{3+}$ and $Eu^{3+}$ ions, which leads to efficient depopulation of the $^5D_4$ excited state of $Tb^{3+}$. The thermally induced variation in the relative emission intensities of these ions results in a noticeable change in the color of the emitted light. This effect is confirmed by the shift of the CIE1931 chromaticity coordinates, from (x, y) = (0.5006, 0.4753) at 93 K to (0.5435, 0.4389) at 373 K (Figure 3b). The integrated emission intensity of $Tb^{3+}$ ions decreases with increasing temperature, reaching approximately 70% of its initial value at 193 K, followed by a further decrease at a significantly lower rate above this temperature (Figure 3c). In contrast, the luminescence intensity of $Eu^{3+}$ ions initially increases with temperature, reaching a maximum corresponding to 118% of its initial value at approximately 193 K. Above this temperature, the $Eu^{3+}$ emission intensity gradually decreases, reaching about 95% of its initial value at 373 K. The initial thermal enhancement of $Eu^{3+}$ emission originates from the growing population of the $Eu^{3+}$ $^5D_0$ emitting level through energy transfer from the $^5D_4$ excited state of $Tb^{3+}$ ions. The correlation between the temperature at which the $Eu^{3+}$ emission begins to decrease and the temperature at which the quenching rate of $Tb^{3+}$ emission changes indicates that, above 193 K, nonradiative depopulation processes of both emitting states become dominant over the thermally enhanced $Tb^{3+}\rightarrow Eu^{3+}$ energy transfer. The difference in the thermal behavior of the luminescence signals of these two ions can be successfully exploited for ratiometric luminescence thermometry by defining the luminescence intensity ratio ($LIR_2$) as follows:

$$LIR_2 = \frac{\int_{600nm}^{725nm} I(Eu^{3+})d\lambda}{\int_{450nm}^{575nm} I(Tb^{3+})d\lambda} \tag{2}$$

As shown in Figure 3d, $LIR_2$ increases rapidly with increasing temperature up to approximately 200 K. Further temperature grow results in less pronounced changes in $LIR_2$, eventually leading to saturation above 300 K. To quantitatively evaluate the thermometric performance of the investigated system, the relative thermal sensitivity was calculated as follows:

$$S_R = \frac{1}{LIR}\frac{\Delta LIR}{\Delta T}\cdot 100\% \qquad (3)$$

The highest sensitivity value, $S_R$ = 0.62% $K^{-1}$, was obtained at 100 K (Figure 3e). Increasing temperature resulted in a gradual decrease in $S_R$ up to approximately 250 K, above which the sensitivity approached values close to zero. These results clearly demonstrate that the thermally induced spectroscopic changes of Gd-BTC:$Tb^{3+}$, $Eu^{3+}$ enable effective temperature sensing in the temperature range below 250 K.

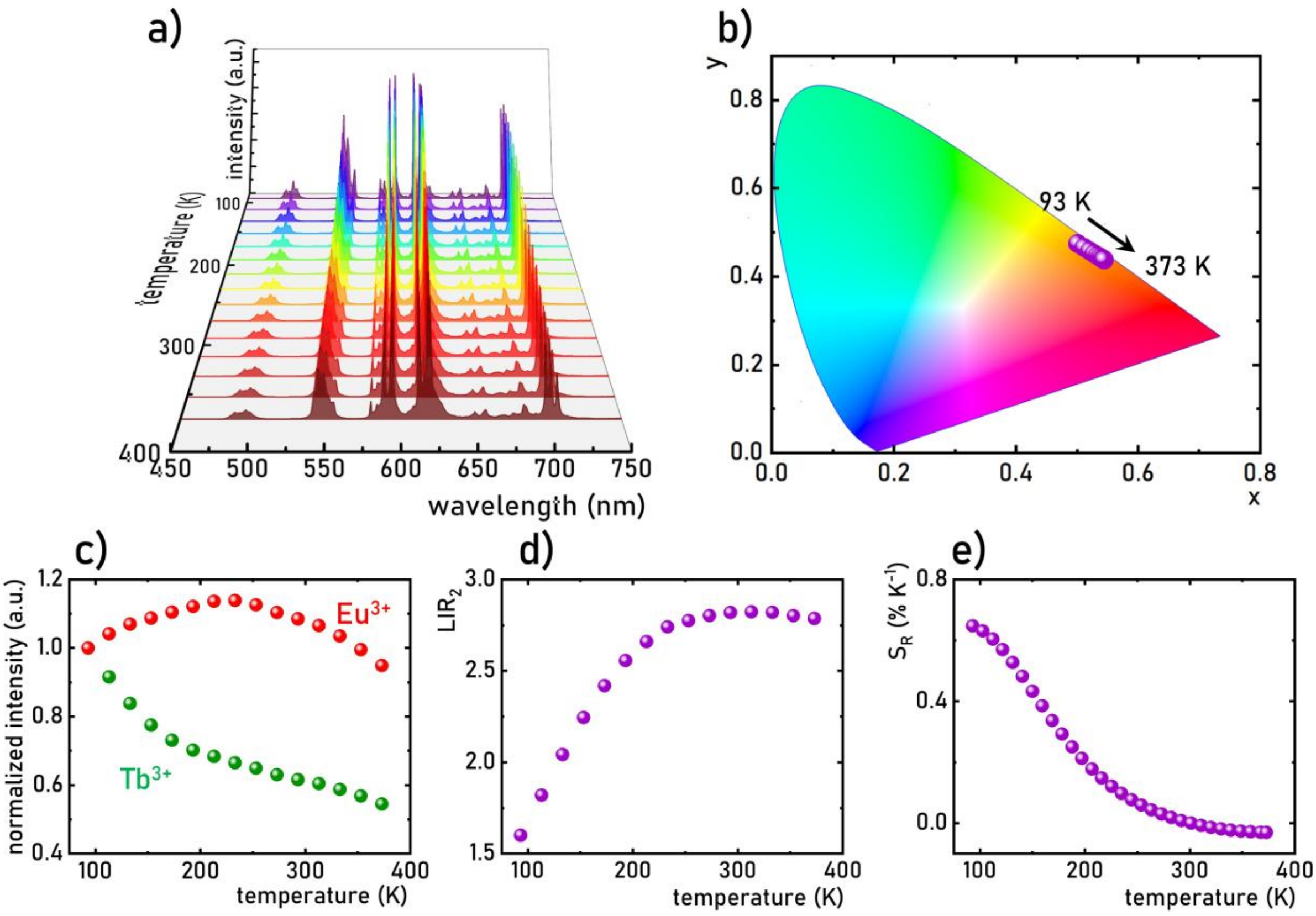

**Figure 3**. Thermal evolution of emission spectra of Gd-BTC:$Tb^{3+}$, $Eu^{3+}$ - a); the influence of the temperature on the CIE1931 chromatic coordinates for this phosphor – b); the influence of the temperature on the integral emission intensity of $Eu^{3+}$ and $Tb^{3+}$ ions (normalized to the emission intensity at 93 K) – c); thermal dependence of $LIR_2$ – d) and corresponding $S_R$ – e).

Although the spectroscopic properties of $Ln^{3+}$ ions are generally considered to be only weakly dependent on their local environment, in MOF-based systems co-doped with more than one lanthanide ion, the nature of the organic linker can influence their luminescence behavior in two distinct ways. First, the MOF structure determines the spatial distribution of $Ln^{3+}$ ions within the framework and, consequently, their mutual separation distances.[32,33] This directly affects both the efficiency and kinetics of interionic energy transfer processes. Second, as in $Tb^{3+}$, $Eu^{3+}$-based systems, the organic ligand acts as both an absorber of excitation energy and a sensitizer for the lanthanide ions through the sequential energy-transfer pathway $S_1 \rightarrow T_1 \rightarrow$ $Ln^{3+}$. Therefore, the energy of the ligand triplet state ($T_1$) plays a crucial role in governing the spectroscopic properties of $Ln^{3+}$-based MOFs. A direct illustration of these effects is provided by the comparison of the luminescence spectra of Gd-BTC:$Eu^{3+}$,$Tb^{3+}$ and Zr-BTC:$Eu^{3+}$,$Tb^{3+}$ recorded at 93 K (Figure 4a). Although both materials exhibit emission bands at nearly identical spectral positions, several important differences can be observed. Most notably, the emission bands in the spectrum of Zr-BTC:$Eu^{3+}$,$Tb^{3+}$ are considerably broadened, resulting in less distinct resolution of the individual Stark components. Furthermore, in Gd-BTC:$Eu^{3+}$,$Tb^{3+}$ the intensity of the emission band associated with $^5D_0 \rightarrow ^7F_1$ electronic transition of $Eu^{3+}$ ions exceeds that of the $Tb^{3+}$ $^5D_4 \rightarrow ^7F_5$ transition, whereas the opposite relationship is observed for Zr-BTC:$Eu^{3+}$,$Tb^{3+}$. Considering that literature reports indicate similar $T_1$ energies for both BTC-based structures, the origin of this difference is more likely associated with the relative spatial distribution of the lanthanide ions. In Gd-BTC, $Ln^{3+}$ ions substitute $Gd^{3+}$ ions and therefore occupy framework node positions, resulting in an average $Ln^{3+}$-$Ln^{3+}$ distance of approximately 4.753 Å.[34] In contrast, in Zr-BTC lanthanide ions can be located at interstitial or defect-related sites or can replace Zr ions. Since Zr in this MOF creates cluster two types of $Ln^{3+}$-$Ln^{3+}$ distances can be considered- one within the cluster (~3.5A) or between clusters ~ 7.34 A. Since, relatively low $Eu^{3+}$ and $Tb^{3+}$ ions concentration is used in this study the energy transfer between ions located withing different clusters should be considered. Therefore, the later distance should be taken into account in this analysis. The shorter $Ln^{3+}$-$Ln^{3+}$ distance in Gd-BTC significantly facilitates $Tb^{3+}$→$Eu^{3+}$ energy transfer, leading to a pronounced reduction of $Tb^{3+}$ emission intensity relative to $Eu^{3+}$ emission already at 93 K. Moreover, the non-nodal localization of lanthanide ions in Zr-BTC:$Eu^{3+}$,$Tb^{3+}$ results in a less well-defined coordination environment, which contributes to the observed broadening of the Stark components. Upon increasing temperature, the luminescence intensity of $Tb^{3+}$ ions in Zr-BTC:$Eu^{3+}$,$Tb^{3+}$ decreases to approximately 40% of its initial value at 290 K, similarly to the behavior observed for Gd-BTC:$Eu^{3+}$,$Tb^{3+}$ (Figure 4c; emission spectra as a function of temperature are available in SI in Figures S9 and S10)). However, unlike in Gd-BTC:$Eu^{3+}$,$Tb^{3+}$, where the $Eu^{3+}$ emission intensity increases with temperature, the $Eu^{3+}$ emission in Zr-BTC:$Eu^{3+}$,$Tb^{3+}$ also decreases upon heating, although at a significantly slower rate than that of $Tb^{3+}$. The more rapid thermal quenching of the $Tb^{3+}$ emission is attributed to thermally activated $Tb^{3+}$→$Eu^{3+}$ energy transfer. The resulting difference in the temperature dependence of the two emission signals can be exploited for ratiometric luminescence thermometry. Accordingly, the $LIR_2$ decreases systematically from approximately 0.90 at 93 K to about 0.55 at 290 K. This pronounced thermal variation of $LIR_2$ enables Zr-BTC:$Eu^{3+}$,$Tb^{3+}$ to operate as a luminescent thermometer, yielding a maximum relative sensitivity of $S_R$ = 0.15% $K^{-1}$ at 300 K (Figure 4e).

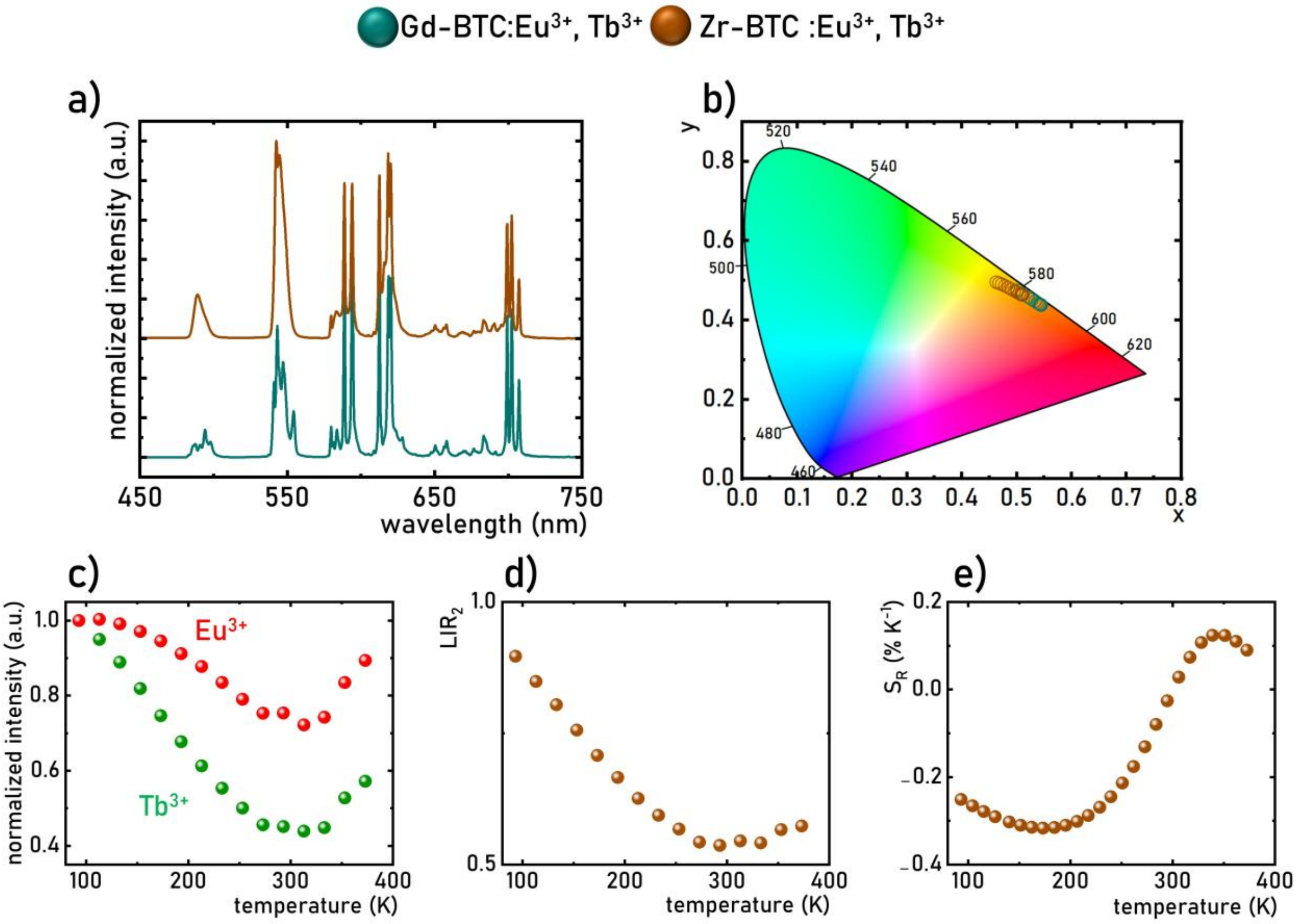

**Figure 4**. Comparison of emission spectra of Eu-BTC and Gd-BTC: $Tb^{3+}$, $Eu^{3+}$ and Zr-BTC: $Tb^{3+}$, $Eu^{3+}$ measured at 93K – a); and corresponding thermal evolutions of CIE1931 chromatic coordinates for these phosphors – b); the influence of the temperature on the integral emission intensities of $Tb^{3+}$ and $Eu^{3+}$ ions in Zr-BTC: $Tb^{3+}$, $Eu^{3+}$ - c); thermal evolution of $LIR_2$ for Zr-BTC: $Tb^{3+}$, $Eu^{3+}$ - d); and corresponding $S_R$ – e).

To evaluate the influence of the ligand on the spectroscopic and thermometric properties of the Zr-based MOFs, emission (Figure 5a) and excitation (Figure 5b) spectra were recorded at 93 K for Zr-BTC:$Tb^{3+}$, $Eu^{3+}$, Zr-BDC:$Tb^{3+}$, $Eu^{3+}$ and Zr-TA:$Tb^{3+}$, $Eu^{3+}$ compared with data obtained for Gd-BTC. Although the spectral positions of the $Tb^{3+}$ and $Eu^{3+}$ emission bands remain essentially unchanged for all investigated materials, significant differences are observed in both the relative intensities of the $Eu^{3+}$ $^5D_0 \rightarrow ^7F_1$ and $^5D_0 \rightarrow ^7F_2$ transitions and the $Tb^{3+}$/$Eu^{3+}$ emission intensity ratio, indicating a pronounced influence of the ligand environment on the luminescence characteristics. Among the investigated materials, Zr-BDC: $Tb^{3+}$, $Eu^{3+}$ exhibits emission dominated by $Tb^{3+}$ ions, whereas the $Eu^{3+}$ emission is relatively weak. A similar, although less pronounced, behavior is observed for Zr-TA:$Tb^{3+}$, $Eu^{3+}$, where $Tb^{3+}$ emission also constitutes the dominant contribution to the luminescence spectrum. In contrast, the emission spectrum of Zr-BTC:$Tb^{3+}$, $Eu^{3+}$ is dominated by $Eu^{3+}$ luminescence. Considering that the $Eu^{3+}$ $^5D_0$ emitting level is populated through a phonon-assisted $Tb^{3+}$→$Eu^{3+}$ energy transfer process, the efficiency of this mechanism would be expected to increase with decreasing average interionic distance. Consequently, shorter metal-metal separations should enhance the depopulation of the $Tb^{3+}$ $^5D_4$ level while increasing the population of the $Eu^{3+}$ $^5D_0$ level, resulting in progressively redder emission. Taking into account that average $Ln^{3+}$-$Ln^{3+}$ distances in the analyzed MOFs changes as follows: Zr-BDC: $Tb^{3+}$, $Eu^{3+}$ (11.7 A)>Zr-TA: $Tb^{3+}$, $Eu^{3+}$ (9.38 A)>Zr-BTC: $Tb^{3+}$, $Eu^{3+}$ (7.34 A)>Gd-BTC: $Tb^{3+}$, $Eu^{3+}$ (4.753 A) observed changes in the emission spectra confirms this hypothesis. Importantly, the proposed mechanism is also reflected in the $Tb^{3+}$/$Eu^{3+}$ luminescence intensity ratio, which reaches its highest value for Zr-BDC: $Tb^{3+}$, $Eu^{3+}$ and its lowest value for Gd-BTC: $Tb^{3+}$, $Eu^{3+}$. Although the emission color is an important characteristic of luminescent materials, the photoluminescence quantum efficiency is another key parameter determining their practical performance. The measured quantum efficiencies amounted to 16.067% for Zr-BTC: $Tb^{3+}$, $Eu^{3+}$, 14.689% for Zr-BDC: $Tb^{3+}$, $Eu^{3+}$ , 14.689% for Zr-TA: $Tb^{3+}$, $Eu^{3+}$ and 22.14 % for Gd-BTC: $Tb^{3+}$, $Eu^{3+}$. To evaluate the applicability of the investigated phosphors for luminescence thermometry, temperature-dependent emission spectra were recorded over the 93-413 K temperature range. The integrated $Tb^{3+}$ emission intensity decreases monotonically with increasing temperature for all investigated samples. The most pronounced thermal quenching below 300 K is observed for Zr-BTC: $Tb^{3+}$, $Eu^{3+}$. However, further temperature increase above this region results in only minor additional changes in the $Tb^{3+}$ emission intensity for this material. In contrast, the $Eu^{3+}$ emission exhibits very similar thermal behavior for all samples. The different temperature dependences of the $Tb^{3+}$ and $Eu^{3+}$ emissions enable the calculation of the luminescence intensity ratio - $LIR_2$, based on Eq.2.

The temperature dependence of $LIR_2$ reveals distinct thermometric characteristics among the investigated materials. For Zr-TA: $Tb^{3+}$, $Eu^{3+}$, $LIR_2$ increases slightly with temperature up to approximately 323 K, above which it begins to decrease. In the case of Zr-BDC: $Tb^{3+}$, $Eu^{3+}$, $LIR_2$ remains nearly constant up to approximately 323 K. For the remaining samples, increasing temperature leads to a gradual decrease in $LIR_2$, with the most pronounced variation observed for Zr-BTC: $Tb^{3+}$, $Eu^{3+}$. As a consequence, Zr-BTC: $Tb^{3+}$, $Eu^{3+}$ exhibits the highest relative thermal sensitivity below 300 K, reaching approximately 0.31% $K^{-1}$. At higher temperatures, the decrease in $LIR_2$ observed for Zr-TA: $Tb^{3+}$, $Eu^{3+}$ above 323 K results in the highest recorded sensitivity of approximately 0.58% $K^{-1}$. Although these values confirm the thermometric capability of the investigated MOFs, their performance remains relatively modest compared with state-of-the-art ratiometric luminescent thermometers reported in the literature.

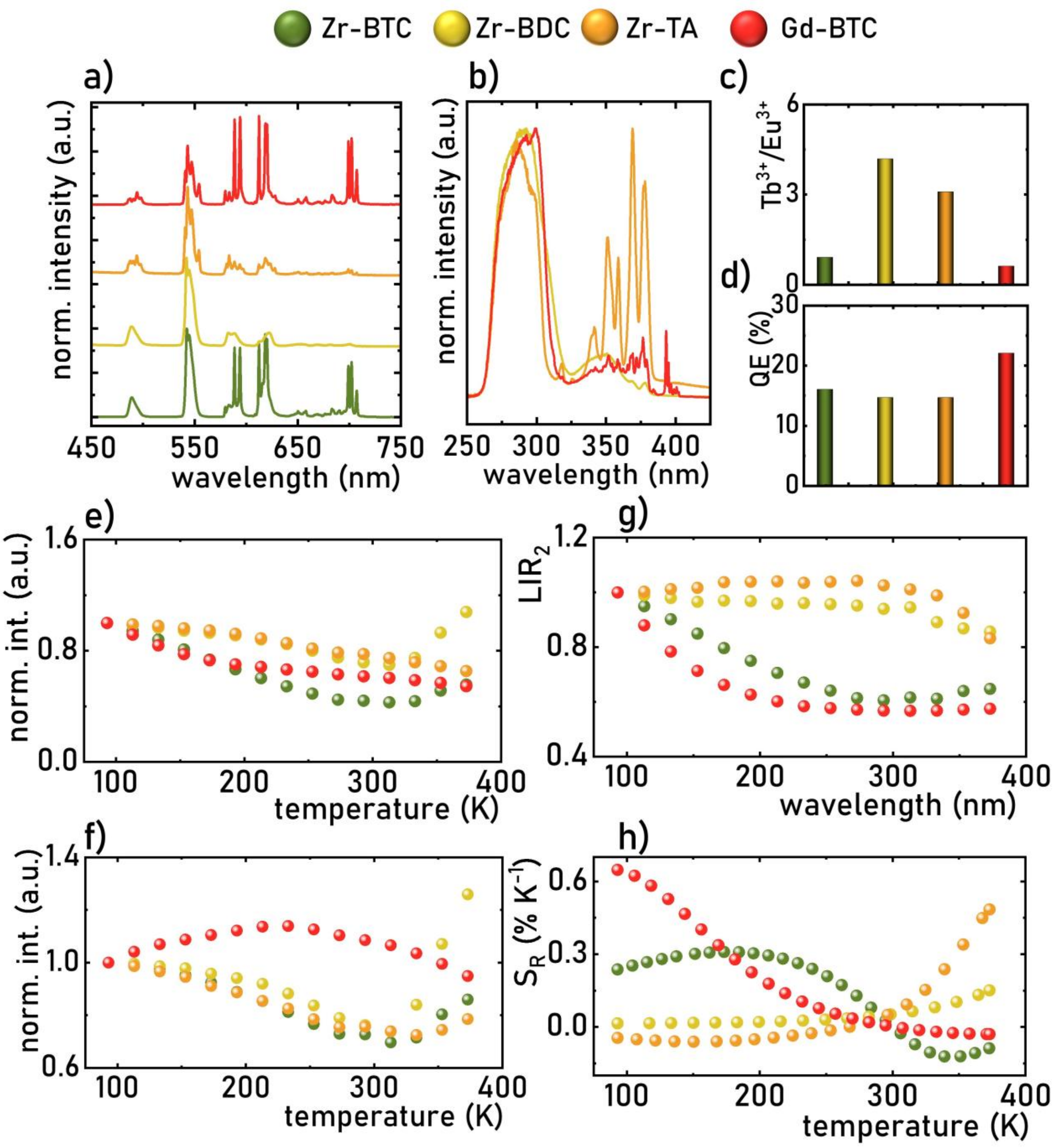

**Figure 5**. Comparison of emission – a) and excitation – b) spectra of Zr-BTC:$Tb^{3+}$, $Eu^{3+}$, Zr-BDC:$Tb^{3+}$, $Eu^{3+}$ , Zr-TA:$Tb^{3+}$, $Eu^{3+}$ measured at 93K; and corresponding $Tb^{3+}$ to $Eu^{3+}$ emission intensity ratio for these phosphors – c); thermal evolution of integrated emission intensity of $Tb^{3+}$ ions – d) and $Eu^{3+}$ ions – f); $LIR_2$ -g) and $S_R$ – h).

The relative changes in the emission intensities of $Tb^{3+}$ and $Eu^{3+}$ ions observed in the investigated samples result in distinct variations in the color of the emitted light. This effect is particularly evident from the CIE 1931 chromaticity coordinates determined at 93 K (Figure 6a). Whereas Zr-TA:$Tb^{3+}$, $Eu^{3+}$ and Zr-BDC:$Tb^{3+}$, $Eu^{3+}$ exhibit emission located in the green region of the chromaticity diagram, the emission of Zr-BTC:$Tb^{3+}$, $Eu^{3+}$ shifts toward the yellow region. Considering the energy-transfer pathway discussed above ($T \rightarrow Tb^{3+} \rightarrow Eu^{3+}$), these variations in the emission color originate from differences in the efficiency of the $Tb^{3+} \rightarrow Eu^{3+}$ energy transfer.[35,36] The probability of this process is governed primarily by two factors: the energy mismatch between the interacting electronic states and the spatial separation between the lanthanide ions. Since the energies of the 4*f* electronic levels of lanthanide ions are only weakly affected by the coordination environment, the ligand framework is expected to play a negligible role in modifying the energetic resonance conditions. Consequently, in the investigated MOFs, the interionic $Tb^{3+}$-$Eu^{3+}$ separation should be the dominant factor controlling the energy-transfer efficiency. Analysis of the shortest metal-metal distances within the investigated frameworks reveals a clear monotonic relationship with the chromaticity coordinates. As the metal-metal distance increases, the $y$ coordinate systematically increases (Figure 6b), whereas the $x$ coordinate decreases (Figure 6c). This trend indicates that shorter interionic distances are associated with more red emission. Described behavior can be readily understood in terms of the enhanced energy-transfer efficiency associated with decreasing interionic distance. As the spatial separation between the $Tb^{3+}$ and $Eu^{3+}$ ions decreases, the $Tb^{3+}$ $^5D_4$ level is more efficiently depopulated through energy transfer to the $Eu^{3+}$ $^5D_0$ level, thereby progressively reducing the relative contribution of the $Tb^{3+}$ emission bands to the overall luminescence spectrum of the phosphor. This trend can be exploited to tailor the emission color of the phosphor by controlling the interionic distance, thereby providing considerable flexibility in tuning its optical properties.

Increasing the temperature induces additional changes in the $x$ and $y$ chromaticity coordinates (Figure 6d and g), which can be quantified by calculating the relative thermal sensitivities, $S_{R,x}$ and $S_{R,y}$ (Figure 6f and Figure S11). The obtained results indicate that below 200 K the highest sensitivities are observed for Zr-BTC:$Tb^{3+}$, $Eu^{3+}$. At higher temperatures, however, the largest values are obtained for Zr-TA:$Tb^{3+}$, $Eu^{3+}$ in terms of the $x$ coordinate. Considering the entire investigated temperature range, the maximum sensitivities reach $S_{R,x}$ = 0.10% $K^{-1}$ for Zr-TA:$Tb^{3+}$, $Eu^{3+}$ (Figure 6f) and $S_{R,y}$ = 0.043% $K^{-1}$ for Zr-BTC:$Tb^{3+}$, $Eu^{3+}$ (Figure 6h). Although these results confirm the temperature dependence of the chromaticity coordinates, the achieved sensitivities remain relatively modest compared with those reported for state-of-the-art luminescent thermometers employing CIE chromaticity coordinates as the temperature readout parameter.[37–40] The magnitude of the temperature-induced variation in the emission color can be quantitatively evaluated by determining the color difference (*CD*) according to the following equation:

$$CD = \sqrt{(x_f - x_i)^2 + (y_f - y_i)^2} \qquad (4)$$

where $x_f$, $y_f$ and $x_i$, $y_i$ represent the final (high temperature) and the initial (low temperature) chcomatic coordinates. The largest variation in *CD* is observed for Gd-BTC, whereas its magnitude decreases rapidly with increasing metal-metal (*M-M*) distance. Importantly, the shape of the trend presented in Figure 6i directly indicates that the *CD* can be systematically tuned by controlling the *M-M* distance. Moreover, the observed relationship suggests that further shortening of the *M-M* distance below that characteristic of Gd-BTC would most likely not result in a substantial additional enhancement of the *CD*. Nevertheless, this interpretation should currently be regarded as a hypothesis and requires experimental verification in future studies.

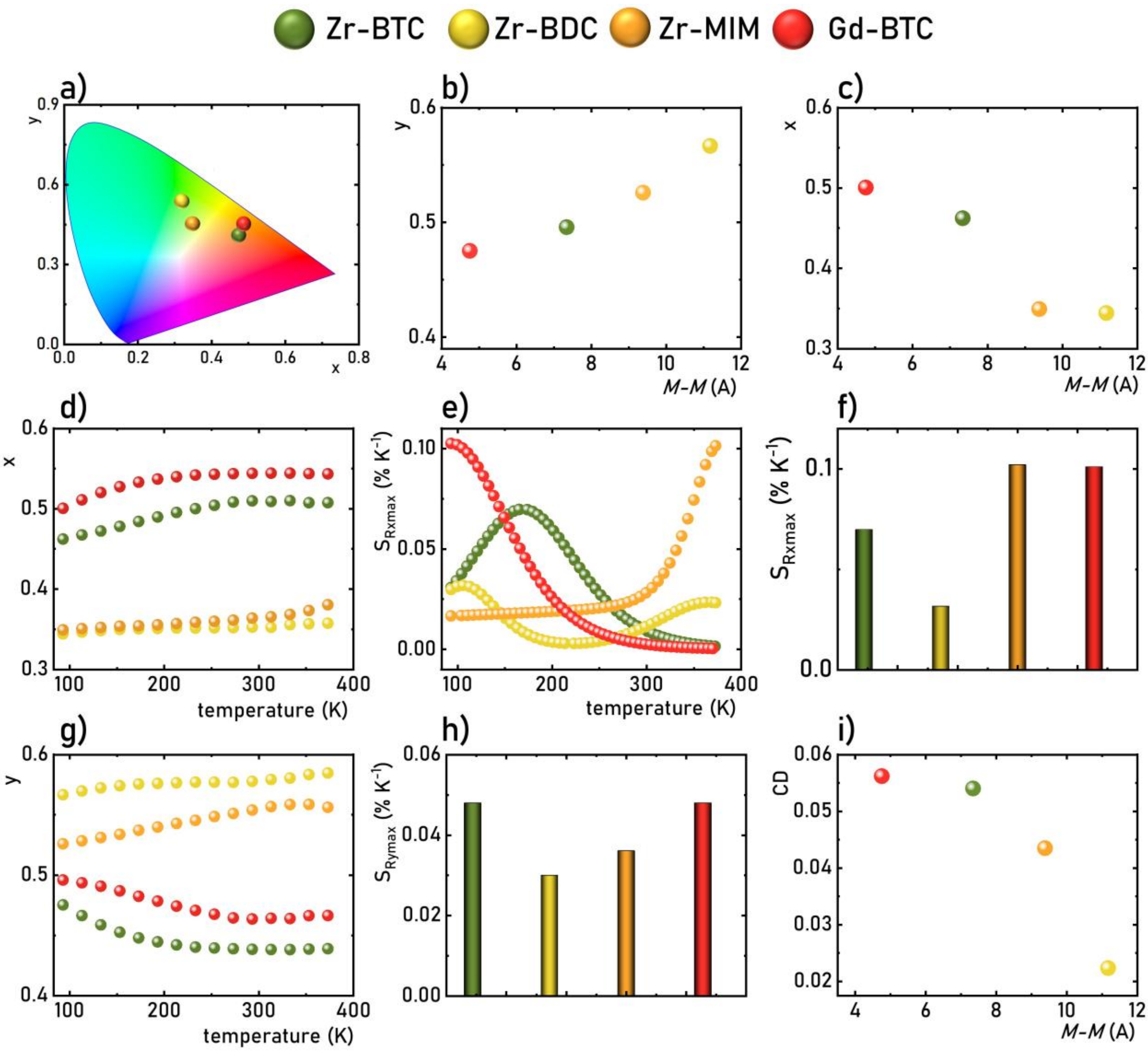

**Figure 6**. CIE 1931 coordinates for different phosphors – a); the influence of metal-metal distance (*M-M*) on the *x* – b) and *y* – c) coordinates obtained at 93K for different Zr-based MOFs; thermal dependence of *x* – d) and *y* – g) chromatic coordinates for these compounds; and corresponding $S_{Rx}$ – e); the comparison of $S_{Rxmax}$ – f) and $S_{Rymax}$ – h); *CD* as a function of *M-M* – i) for these compounds.

As demonstrated in the previous section, temperature induces only relatively small variations in the CIE 1931 chromaticity coordinates of the investigated MOFs. Nevertheless, an alternative and more practical approach to colorimetric analysis is based on the direct evaluation of luminescence photographs. Unlike the conventional CIE analysis, which requires the acquisition of emission spectra and therefore the use of spectroscopic detectors to determine the chromaticity coordinates, image-based analysis relies solely on the signals independently recorded in the red (R), green (G), and blue (B) channels of a standard digital camera.[37,38] By analyzing the intensity ratios between these channels, rapid two-dimensional thermal imaging can be performed in a simple, inexpensive, and user-friendly manner. Therefore, the applicability of this approach was evaluated for the investigated MOF systems. For this purpose, a series of luminescence photographs was recorded as a function of temperature. The obtained images reveal distinct temperature-dependent variations in the emission color (Figure 7a, Figures S12-S18). Depending on whether the emission was dominated by $Tb^{3+}$ or $Eu^{3+}$ ions, the observed luminescence ranged from green to red. These differences are particularly evident when comparing Zr-BDC:$Tb^{3+}$, $Eu^{3+}$, which exhibits intense green emission, with Gd-BTC:$Tb^{3+}$, $Eu^{3+}$, characterized by pronounced yellow luminescence. In contrast, samples such as Gd-BTC:$Tb^{3+}$, $Eu^{3+}$ and Zr-BTC:$Tb^{3+}$, $Eu^{3+}$ display yellow emission originating from the superposition of green $Tb^{3+}$ and red $Eu^{3+}$ emissions of comparable intensity. The manner in which this luminescence is registered by the RGB channels of a digital camera can be understood by comparing the representative emission spectrum of Zr-TA:$Tb^{3+}$, $Eu^{3+}$ with the spectral sensitivity profiles of the camera channels (Figure 7b). This comparison clearly demonstrates that the red channel records almost exclusively the $Eu^{3+}$ emission, with only a minor contribution from the $Tb^{3+}$ $^5D_4 \rightarrow {}^7F_4$ transition. In contrast, the green channel detects the entire $Tb^{3+}$ emission together with the $Eu^{3+}$ $^5D_0 \rightarrow {}^7F_1$ transition and a partial contribution from the $^5D_0 \rightarrow {}^7F_2$ transition. Meanwhile, the blue channel records almost exclusively the $Tb^{3+}$ $^5D_4 \rightarrow {}^7F_5$ emission. Consequently, decomposition of the luminescence photographs into their individual RGB channels provides an alternative way of monitoring thermally induced changes in the emission characteristics. This behavior is illustrated for the representative Zr-TA:$Tb^{3+}$, $Eu^{3+}$ sample (Figure 7c). Whereas only minor intensity changes are observed in the red channel with increasing temperature, the blue channel exhibits a pronounced thermal quenching of the emission. Considering that all investigated materials exhibit thermally induced changes in the relative emission intensities of green-emitting $Tb^{3+}$ and red-emitting $Eu^{3+}$ ions, the most intuitive choice for ratiometric analysis would be the *R/G* intensity ratio (Figure 7d). Surprisingly, the experimental results reveal that this parameter does not exhibit a significant monotonic dependence on temperature for most of the investigated samples. This behavior can be readily explained by the broad spectral response of the green channel, which simultaneously collects a substantial fraction of both $Tb^{3+}$ and $Eu^{3+}$ emissions. As a consequence, the thermal quenching of the $Tb^{3+}$ emission is partially compensated by the increasing relative contribution of $Eu^{3+}$ emission, thereby suppressing the overall temperature dependence of the *R/G* ratio. A considerably more effective, although less intuitive, parameter is the *B/G* intensity ratio. In this case, a nearly monotonic temperature dependence is observed for almost all investigated samples (Figure 7e). While the *B/G* ratio remains nearly constant for Zr-BTC:$Tb^{3+}$, $Eu^{3+}$ and Zr-BDC:$Tb^{3+}$, $Eu^{3+}$, a pronounced thermal decrease is observed for the remaining materials, indicating that the $Tb^{3+}$ emission undergoes significantly stronger thermal quenching than the $Eu^{3+}$ emission. The most pronounced variation is found for Zr-TA:$Tb^{3+}$, $Eu^{3+}$, for which the *B/G* ratio decreases by more than 50% over the investigated temperature range. However, this sample exhibits only negligible changes above 300 K. To quantitatively evaluate the thermometric performance, the relative thermal sensitivity was calculated (Figure 7f). Below 300 K, the highest sensitivity values, exceeding 0.30% $K^{-1}$, were obtained for Zr-TA:$Tb^{3+}$, $Eu^{3+}$. A comparison of the room-temperature sensitivities reveals the following trend: Gd-BTC:$Tb^{3+}$, $Eu^{3+}$ < Zr-TA:$Tb^{3+}$, $Eu^{3+}$ (Figure 7g). Because RGB image-based temperature readout has only recently emerged as a thermometric strategy, and only a limited number of studies have reported relative sensitivities determined using this approach, a comprehensive comparison of the thermometric performance of the investigated MOFs with previously reported luminescent thermometers remains difficult. Nevertheless, an important conclusion arising from the present study is that, within this image-based readout methodology, Zr-TA:$Tb^{3+}$, $Eu^{3+}$ exhibits considerably higher thermal sensitivity than the widely studied Gd-BTC:$Tb^{3+}$, $Eu^{3+}$ reference material, highlighting its strong potential for practical camera-based luminescence thermometry.

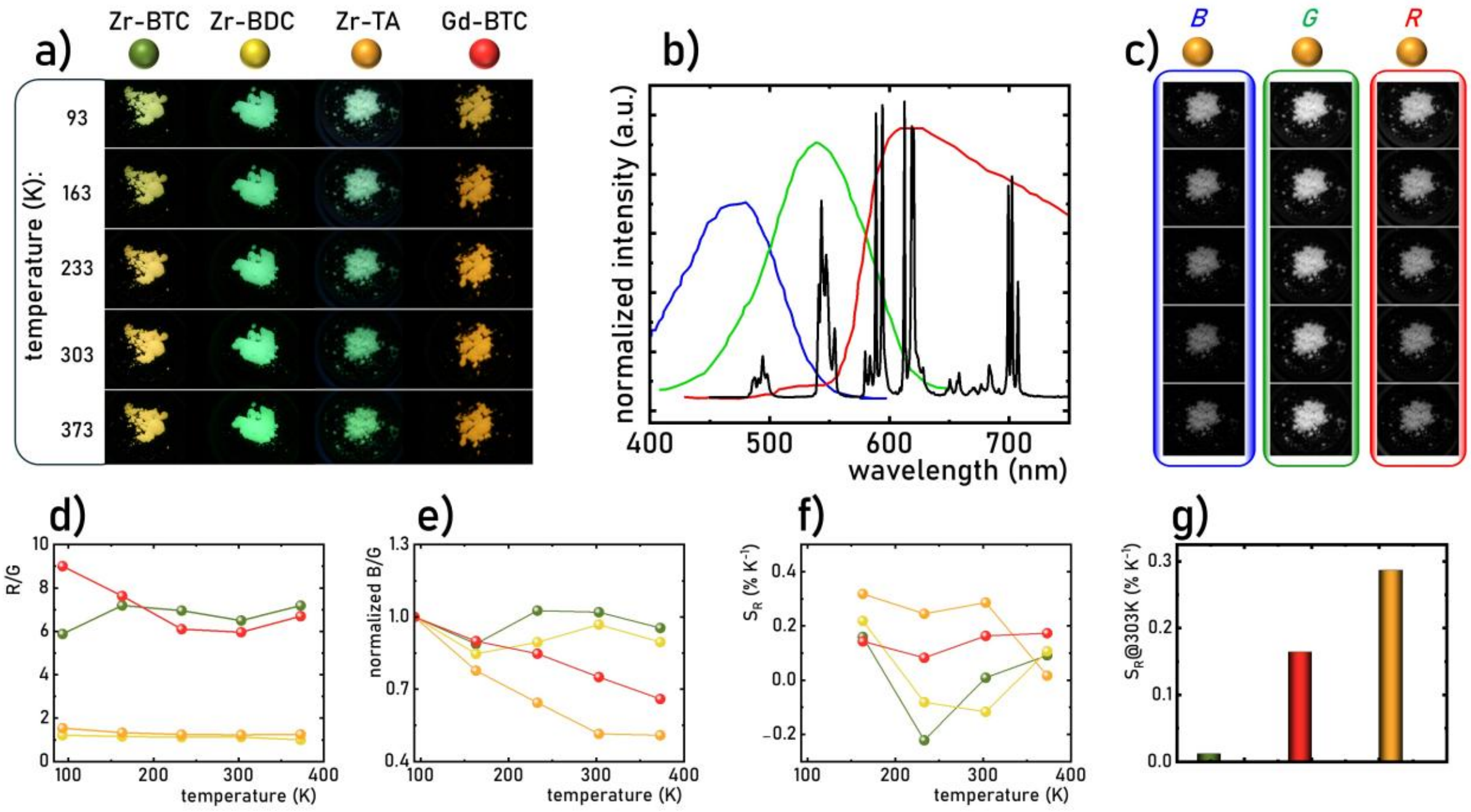

**Figure 7**. Photos of luminescence from different compounds at different temperatures – a); comparison of emission spectra of Zr-BTC:$Tb^{3+}$,$Eu^{3+}$ measured at 93K with sensitivities curves of blue (B), green (G) and red (R) channels of digital camera – b), intensities maps of luminescence of Zr-TA:$Tb^{3+}$,$Eu^{3+}$ captured at B, G and R channels – c); thermal dependence of *R/B* for different phosphors – d); thermal dependence of *B/G* for different phosphors – e); and corresponding $S_R$ – f); comparison of $S_R$ at 300K for different phosphors – g).

## Conclusions

In this work, the spectroscopic properties of Gd-BTC:$Tb^{3+}$,$Eu^{3+}$, Zr-BTC:$Tb^{3+}$,$Eu^{3+}$, Zr-BDC:$Tb^{3+}$,$Eu^{3+}$ and Zr-TA:$Tb^{3+}$,$Eu^{3+}$ were investigated to evaluate their potential for luminescent thermometry. To develop a ratiometric luminescent thermometer with the capability of direct visual readout, $Tb^{3+}$ and $Eu^{3+}$ ions were introduced as co-dopants. In such systems, the $T_1 \rightarrow Tb^{3+} \rightarrow Eu^{3+}$ energy transfer pathway exhibits a strong temperature dependence due to the differences in the energies of the excited states of the two lanthanide ions. As demonstrated in this study, the presence of $Gd^{3+}$ ions in Gd-BTC:$Tb^{3+}$,$Eu^{3+}$ significantly modifies the ligand-to-lanthanide energy transfer process. Therefore, replacing $Gd^{3+}$ with $Zr^{4+}$ was expected to eliminate this effect and provide greater control over the energy transfer dynamics. Although Gd-BTC:$Tb^{3+}$,$Eu^{3+}$ is the metal-organic framework most commonly employed for this purpose, the present study demonstrates that an appropriate selection of the organic ligand enables effective tuning of the thermometric performance of Zr-based MOFs. It was shown that the choice of the MOF structure has a pronounced influence on the emission color through modification of the metal-metal distance. Specifically, an shortening in the metal-metal separation resulted in a higher relative intensity of $Tb^{3+}$ emission with respect to $Eu^{3+}$ emission. This behavior was attributed to the facilitated depopulation of the $^5D_4$ state of $Tb^{3+}$ ions to $^5D_0$ state of $Eu^{3+}$ ions via $Eu^{3+}$ to $Tb^{3+}$ energy transfer. Moreover, the excellent correlation between the CIE 1931 chromaticity coordinates recorded at low temperature, color difference and the metal-metal distance demonstrates that the emission color can be systematically tailored by controlling this structural parameter. The investigated materials were demonstrated to be suitable for luminescent thermometry using both the conventional ratiometric approach and the analysis of CIE 1931 chromaticity coordinates. However, in both cases the relative sensitivity values were only moderate. To overcome this limitation, a new approach for MOF-based luminescent thermometry was proposed, relying on the analysis of the RGB channels of a conventional digital camera. The analysis revealed that the red (R) channel records almost exclusively the emission of $Eu^{3+}$ ions, whereas the green (G) channel contains contributions from both $Tb^{3+}$ and $Eu^{3+}$ emissions. Consequently, the blue (B) channel, which selectively records the $Tb^{3+}$ emission, was employed instead of the green channel. This strategy proved to be highly effective for the Zr-TA:$Tb^{3+}$,$Eu^{3+}$ systems, demonstrating their suitability for temperature sensing applications. The highest relative thermal sensitivity, reaching 0.28% $K^{-1}$, was obtained for Zr-TA:$Tb^{3+}$,$Eu^{3+}$. A major advantage of the proposed methodology is the use of a conventional digital camera as the only detection device, providing a simple, rapid, and low-cost alternative to spectroscopic instrumentation. Furthermore, this approach enables not only point temperature sensing but also spatially resolved thermal imaging, considerably expanding the application potential of these luminescent MOF materials.

**Acknowledgements**

The authors would like to acknowledge support within the Joint Mobility Projects for years 2024 - 2026 between Polish Academy of Sciences and Vietnam Academy of Science and Technology. This work was supported by the Vietnam Academy of Science and Technology under project No. QTPL01.02/24-25. The authors would like to express their sincere gratitude to Dr. Anna Świderska-Środa for the FTIR measurements, Mr. Jan Mizeracki for the SEM characterization, and Dr. Agnieszka Opalińska and Prof. Witold Lojkowski for their valuable assistance.

**References**

(1) Xiuqin, O.; Lin, P.; Haichen, G.; Yichen, W.; Jianwei, L. Temperature-Dependent Crystallinity and Morphology of LiFePO4 Prepared by Hydrothermal Synthesis. J. Mater. Chem. 2012, 22 (18), 9064–9068. https://doi.org/10.1039/c2jm30191a.

(2) Sholl, D. S.; Lively, R. P. Defects in Metal–Organic Frameworks: Challenge or Opportunity? J. Phys. Chem. Lett. 2015, 6 (17), 3437–3444. https://doi.org/10.1021/acs.jpclett.5b01135.

(3) Scherrer, P. Bestimmung der Grösse und der inneren Struktur von Kristallteilchen mittels Röntgenstrahlen. Nachr. Ges. Wiss. Göttingen, 1918, 98-100.

(4) Fang, Z.; Bueken, B.; De Vos, D. E.; Fischer, R. A. Defect-Engineered Metal–Organic Frameworks. Angew. Chem. Int. Ed. 2015, 54 (25), 7234–7254. https://doi.org/10.1002/anie.201411540.

(5) Vermoortele, F.; Bueken, B.; Le Bars, G.; Van de Voorde, B.; Vandichel, M.; Houthoofd, K.; Vimont, A.; Daturi, M.; Waroquier, M.; Van Speybroeck, V.; Kirschhock, C.; De Vos, D. E. Synthesis Modulation as a Tool To Increase the Catalytic Activity of Metal–Organic Frameworks: The Unique Case of UiO-66(Zr). J. Am. Chem. Soc. 2013, 135 (31), 11465–11468. https://doi.org/10.1021/ja405078u.

(6) Feng, L.; Wang, K.-Y.; Lv, X.-L.; Yan, T.-H.; Zhou, H.-C. Hierarchically Porous Metal–Organic Frameworks: Synthetic Strategies and Applications. Natl. Sci. Rev. 2020, 7 (11), 1743–1758. https://doi.org/10.1093/nsr/nwz170.

(7) Bennett, T. D.; Cheetham, A. K. Amorphous Metal–Organic Frameworks. Acc. Chem. Res. 2014, 47 (5), 1555–1562. https://doi.org/10.1021/ar5000314.

(8) Howarth, A. J.; Liu, Y.; Li, P.; Li, Z.; Wang, T. C.; Hupp, J. T.; Farha, O. K. Chemical, Thermal and Mechanical Stabilities of Metal–Organic Frameworks. Nat. Rev. Mater. 2016, 1 (3), 15018. https://doi.org/10.1038/natrevmats.2015.18.

(9) Kalarakoppa, K. D.; Prabhu, A. N.; Nayak, R.; Prabhu, N.; Saquib, M.; Shetty, S.; Naik, K. A Decoupling Strategy to Optimize Power Density in Flexible Thermoelectric Devices Using a ZIF-67 Doped Polypyrrole Bio Binder-Based Hybrid Ink. Mater. Adv. 2026, 7 (2), 826–844. https://doi.org/10.1039/d5ma00819k.

(10) Robles-Águila, M. J.; Reyes-Avendaño, J. A.; Silva, R.; Bravo-Arredondo, J. M. Green Synthesis of Mn-MOFs with Tunable Optical and Structural Properties Through Ligand Functionality and Framework Design. J. Inorg. Organomet. Polym. Mater. 2026, 36 (3), 1745–1760. https://doi.org/10.1007/s10904-025-03955-6.

(11) Magnetic MOF-808 as a Novel Adsorbent for Toxic Metal Removal from Aqueous Solutions. Environ. Sci. Adv. 2022, 1 (2), 182–191. https://doi.org/10.1039/d2va00010e.

(12) Butova, V. V.; Vetlitsyna-Novikova, K. S.; Pankin, I. A.; Charykov, K. M.; Trigub, A. L.; Soldatov, A. V. Microwave Synthesis and Phase Transition in UiO-66/MIL-140A System.

Microporous Mesoporous Mater. 2020, 296, 109998. https://doi.org/10.1016/j.micromeso.2020.109998.

(13) Mu, Z.; Hu, Y.; Chen, L.; Wang, X.; Chen, R.; Wang, T.; Fu, Y.; Xu, J. Synthesis of Bi3+ and Gd3+ Doped ZnB2O4 for Evaluation as Potential Materials in Luminescent Display Applications. Displays 2014, 35 (3), 147–151. https://doi.org/10.1016/j.displa.2014.04.003.

(14) Pacold, J. I.; Tatum, D. S.; Seidler, G. T.; Raymond, K. N.; Zhang, X.; Stickrath, A. B.; Mortensen, D. R. Direct Observation of 4f Intrashell Excitation in Luminescent Eu Complexes by Time-Resolved X-ray Absorption Near Edge Spectroscopy. J. Am. Chem. Soc. 2014, 136 (11), 4186–4191. https://doi.org/10.1021/ja407924m.

(15) Mara, M. W.; Tatum, D. S.; March, A.-M.; Doumy, G.; Moore, E. G.; Raymond, K. N. Energy Transfer from Antenna Ligand to Europium(III) Followed Using Ultrafast Optical and X-ray Spectroscopy. J. Am. Chem. Soc. 2019, 141 (28), 11071–11081. https://doi.org/10.1021/jacs.9b02792.

(16) Baweja, S.; Kappes, M. M.; Schäfer, A.; Znotins, A.; Zielke, R.; Holzer, C.; Neumann, T.; Kruck, C.; Seitz, M. Direct Measurement of 4f–4f Transitions and Electronic Hot Bands in Lanthanoid-Antenna Complexes by Helium-Tagging Spectroscopy: Toward Molecular-Scale Trapped Ion Qubits. J. Phys. Chem. Lett. 2025, 16 (51), 13046–13053. https://doi.org/10.1021/acs.jpclett.5c03294.

(17) Zhang, X.; Huang, Y.; Gong, M. Dual-Emitting Ce3+, Tb3+ Co-Doped LaOBr Phosphor: Luminescence, Energy Transfer and Ratiometric Temperature Sensing. Chem. Eng. J. 2017, 307, 291–299. https://doi.org/10.1016/j.cej.2016.08.087.

(18) Kang, H.; Peng, J.; Zhang, Z.; Zhou, W. Fluorescent Strengthening Effect of Co-Doped Inert Rare Earth Ions (La3+, Gd3+, Lu3+) on White-Light-Emitting of Eu–Tb(Btc) Coordination Polymers. J. Lumin. 2022, 247, 118904. https://doi.org/10.1016/j.jlumin.2022.118904.

(19) Binnemans, K. Interpretation of Europium(III) Spectra. Coord. Chem. Rev. 2015, 295, 1–45. https://doi.org/10.1016/j.ccr.2015.02.015.

(20) Janulevicius, M.; Marmokas, P.; Misevicius, M.; Grigorjevaite, J.; Mikoliunaite, L.; Sakirzanovas, S.; Katelnikovas, A. Luminescence and Luminescence Quenching of Highly Efficient Y2Mo4O15:Eu3+ Phosphors and Ceramics. Sci. Rep. 2016, 6 (1), 26098. https://doi.org/10.1038/srep26098.

(21) Li, G.; Cai, G.; Wang, X.-J. Local Structure–Luminescence Correlation in Eu3+-Doped Phosphors: A Comprehensive Review. Adv. Photonics Res. 2025, 6 (11), 2500142. https://doi.org/10.1002/adpr.202500142.

(22) He, X.; Wang, J.; Niu, G.; Zhu, D. Eu3+ Functionalized Gd-BTC: Turn-off Fluorescent Switch for Selectively Detecting Acetone and Fe3+. J. Mol. Struct. 2022, 1267, 133663. https://doi.org/10.1016/j.molstruc.2022.133663.

(23) Lucena, M. A. M.; Oliveira, M. F. L.; Arouca, A. M.; Talhavini, M.; Ferreira, E. A.; Alves, S., Jr.; Veiga-Souza, F. H.; Weber, I. T. Application of the Metal–Organic Framework [Eu(BTC)] as a Luminescent Marker for Gunshot Residues: A Synthesis, Characterization, and Toxicity Study. ACS Appl. Mater. Interfaces 2016, 9 (5), 4684–4691. https://doi.org/10.1021/acsami.6b13474.

(24) He, X.; Liu, Y.; Wang, Q.; Wang, T.; He, J.; Peng, A.; Qi, K. Facile Fabrication of Eu-Based Metal–Organic Frameworks for Highly Efficient Capture of Tetracycline Hydrochloride from Aqueous Solutions. Sci. Rep. 2023, 13 (1), 11107. https://doi.org/10.1038/s41598-023-38425-x.

(25) Carneiro Neto, A. N.; Moura, R. T., Jr.; Shyichuk, A.; Paterlini, V.; Piccinelli, F.; Bettinelli, M.; Malta, O. L. Theoretical and Experimental Investigation of the Tb3+ → Eu3+ Energy Transfer Mechanisms in Cubic A3Tb0.90Eu0.10(PO4)3 (A = Sr, Ba) Materials. J. Phys. Chem. C 2020, 124 (18), 10105–10116. https://doi.org/10.1021/acs.jpcc.0c00759.

(26) Bao, G.; Wong, K.-L.; Jin, D.; Tanner, P. A. A Stoichiometric Terbium-Europium Dyad Molecular Thermometer: Energy Transfer Properties. Light Sci. Appl. 2018, 7 (1), 96. https://doi.org/10.1038/s41377-018-0097-7.

(27) Ramakrishna, P.; Padhi, R. K.; Mohapatra, D. K.; Jena, H.; Panigrahi, B. S. Structural Characterization, Gd3+ → Eu3+ Energy Transfer and Radiative Properties of Gd/Eu in Codoped Li2O–ZnO–SrO–B2O3–P2O5 Glass. Opt. Mater. 2022, 125, 112060. https://doi.org/10.1016/j.optmat.2022.112060.

(28) Zaman, F.; Abbas, J.; Ullah, I.; Khan, A.; Saqib, N. U.; Mukamil, S.; Albargi, H. B.; Rooh, G.; Srisittipokakun, N.; Rachniyom, W.; Intachai, N.; Kothan, S.; Kaewkhao, J. Investigation of Energy Transfer Mechanism in Gd3+ to Sm3+ and Eu3+ in Borate Glasses for the Application of Solid-State Lighting Devices. Solid State Sci. 2025, 163, 107878. https://doi.org/10.1016/j.solidstatesciences.2025.107878.

(29) Ngom, F.; Chang, A.; Blais, C.; Daiguebonne, C.; Suffren, Y.; Camara, M.; Calvez, G.; Bernot, K.; Guillou, O. Halogen-Bonds-Based Strategy for the Design of Highly Luminescent Lanthanide Coordination Polymers as Taggants for Plastic Waste Sorting. Inorg. Chem. 2024, 63 (28), 13048–13058. https://doi.org/10.1021/acs.inorgchem.4c01866.

(30) Gálico, D. A.; Murugesu, M. Inside-Out/Outside-In Tunability in Nanosized Lanthanide-Based Molecular Cluster-Aggregates: Modulating the Luminescence Thermometry Performance via Composition Control. ACS Appl. Mater. Interfaces 2021, 13 (39), 47052–47060. https://doi.org/10.1021/acsami.1c13684.

(31) Wang, J.; Suffren, Y.; Daiguebonne, C.; Freslon, S.; Bernot, K.; Calvez, G.; Le Pollès, L.; Roiland, C.; Guillou, O. Multi-Emissive Lanthanide-Based Coordination Polymers for Potential Application as Luminescent Bar-Codes. Inorg. Chem. 2019, 58 (4), 2659–2668. https://doi.org/10.1021/acs.inorgchem.8b03277.

(32) Psalti, A. E.; Andriotou, D.; Diamantis, S. A.; Chatz-Giachia, A.; Pournara, A.; Manos, M. J.; Hatzidimitriou, A.; Lazarides, T. Mixed-Metal and Mixed-Ligand Lanthanide Metal–Organic Frameworks Based on 2,6-Naphthalenedicarboxylate: Thermally Activated Sensitization and White-Light Emission. Inorg. Chem. 2022, 61 (30), 11959–11972. https://doi.org/10.1021/acs.inorgchem.2c01703.

(33) Liu, X.; Akerboom, S.; Jong, M. de; Mutikainen, I.; Tanase, S.; Meijerink, A.; Bouwman, E. Mixed-Lanthanoid Metal–Organic Framework for Ratiometric Cryogenic Temperature Sensing. Inorg. Chem. 2015, 54 (23), 11323–11329. https://doi.org/10.1021/acs.inorgchem.5b01924.

(34) Zeleňák, V.; Almáši, M.; Zeleňáková, A.; Hrubovčák, P.; Tarasenko, R.; Bourelly, S.; Llewellyn, P. Large and Tunable Magnetocaloric Effect in Gadolinium-Organic Framework: Tuning by Solvent Exchange. Sci. Rep. 2019, 9 (1), 15572. https://doi.org/10.1038/s41598-019-51590-2.

(35) Rodrigues, M. O.; Dutra, J. D. L.; Nunes, L. A. O.; de Sá, G. F.; de Azevedo, W. M.; Silva, P.; Paz, F. A. A.; Freire, R. O.; A. Júnior, S. Tb3+→Eu3+ Energy Transfer in Mixed-Lanthanide-Organic Frameworks. J. Phys. Chem. C 2012, 116 (37), 19951–19957. https://doi.org/10.1021/jp3054789.

(36) Wang, B.; Ren, Q.; Hai, O.; Wu, X. Luminescence Properties and Energy Transfer in Tb3+ and Eu3+ Co-Doped Ba2P2O7 Phosphors. RSC Adv. 2017, 7 (25), 15222–15227. https://doi.org/10.1039/c6ra28122b.

(37) Abe, Y.; Szymczak, M.; Zeler, J.; Szukiewicz, R.; Marciniak, L. Interplay of Optical Traps and Vibronic Thermalization Governing Mn2+ Luminescence in Ca19Ce(PO4)14: Implications for Temperature Sensing and Thermal Threshold Indicator. Laser Photonics Rev. n/a (n/a), e71411. https://doi.org/10.1002/lpor.71411.

(38) Marciniak, L.; Szymczak, M. Visual Luminescence Thermometry Enabled by Phase-Transition-Activated Cross Relaxation of Tb3+ Ions. Mater. Horiz. 2026, 13 (14), 7157–7172. https://doi.org/10.1039/d6mh00684a.

(39) Yin, Y.; Jiang, M.; Tian, L. Novel Phosphor GdY2SbO7 Co-Dope with Eu3+ and Bi3+ for Optical Thermometer. Heliyon 2024, 10 (3), e24496. https://doi.org/10.1016/j.heliyon.2024.e24496.

(40) Huang, W.; Li, Z.; Yang, N.; Ye, Z.; Cao, R.; Shi, J.; Wang, J. A Four-Mode Optical Thermometer Designed in Double Perovskite Ca2GdNbO6:Bi3+,Eu3+. J. Alloys Compd. 2024, 1002, 175220. https://doi.org/10.1016/j.jallcom.2024.175220.